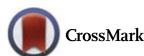

# New Journal of Physics
The open access journal at the forefront of physics

**PAPER**

# Monte Carlo sampling from the quantum state space. I

Jiangwei Shang[1], Yi-Lin Seah[1,2], Hui Khoon Ng[1,3,4], David John Nott[5] and Berthold-Georg Englert[1,2,4]

1. Centre for Quantum Technologies, National University of Singapore, 3 Science Drive 2, Singapore 117543, Singapore
2. Department of Physics, National University of Singapore, 2 Science Drive 3, Singapore 117542, Singapore
3. Yale-NUS College, 6 College Avenue East, Singapore 138614, Singapore
4. MajuLab, CNRS-UNS-NUS-NTU International Joint Research Unit, UMI 3654, Singapore
5. Department of Statistics and Applied Probability, National University of Singapore, 6 Science Drive 2, Singapore 117546, Singapore

E-mail: jiangwei.shang@quantumlah.org and s89@u.nus.edu

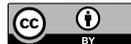

## Abstract

High-quality random samples of quantum states are needed for a variety of tasks in quantum information and quantum computation. Searching the high-dimensional quantum state space for a global maximum of an objective function with many local maxima or evaluating an integral over a region in the quantum state space are but two exemplary applications of many. These tasks can only be performed reliably and efficiently with Monte Carlo methods, which involve good samplings of the parameter space in accordance with the relevant target distribution. We show how the standard strategies of rejection sampling, importance sampling, and Markov-chain sampling can be adapted to this context, where the samples must obey the constraints imposed by the positivity of the statistical operator. For illustration, we generate sample points in the probability space of qubits, qutrits, and qubit pairs, both for tomographically complete and incomplete measurements. We use these samples for various purposes: establish the marginal distribution of the purity; compute the fractional volume of separable two-qubit states; and calculate the size of regions with bounded likelihood.

## 1. Introduction

Many situations in quantum information and quantum computation call for a random sample from the space of quantum states. This can be in the numerical testing of the typicality of entanglement among states from a bipartite quantum system, understanding of the efficacy of a gate implementation or a noise protection scheme by examining its performance on randomly selected states, computation of some quantity of interest over a subset of quantum states, integration over a region of states, optimization of a function on the state space with a complicated landscape, etc. In every case, a quantitative conclusion can be drawn only if one first specifies what the word 'random' means, i.e., according to which distribution we are drawing states, and then have an efficient way of sampling from the state space in accordance with that distribution.

In many cases, in the absence of additional information or when desiring caution against pre-biasing the results, what one means by drawing a random sample translates into sampling from a 'uniform' distribution of states that treats every state 'fairly.' This is certainly an appropriate attitude when dealing with a discrete (sub)set of quantum states $\{\rho_t\}_{t=1}^N$ so that the uniform distribution is simply one that assigns probability $1/N$ to each state $\rho_t$. For a continuous set of states, the notion of a uniform distribution, or more generally, an 'uninformative' distribution [1, 2], is an ill-defined one that depends highly on the choice of parameterization of the state space and the criterion for uniformity. One hence needs to specify the desired distribution and, depending on the choice, how one samples according to that distribution may not be easy or obvious.

One often-used choice of random distribution is defined by writing the state $\rho$ as $U\Lambda U^\dagger$, with $U$ a Haar-random unitary matrix, and $\Lambda$ a diagonal nonnegative matrix with entries chosen according to the Lebesgue measure on real space. [3] and [4] describe how one samples from this distribution in a simple and computationally efficient way, even in high-dimensional problems, and employ such samples for the estimation of the proportion of entangled to separable states in bipartite state spaces. Another popular sampling approach is that introduced in [5], where the distribution of states is defined by the sampling method itself: sample from a rotationally invariant measure defined on the set of pure states in a composite system built from the original system and a duplicate copy, and then trace out the duplicate copy to arrive at a mixed state of the system, drawn from the thus-defined rotationally symmetric distribution. Later works, including those of [6] and [7] followed the same idea, but employed Monte Carlo techniques to help with the sampling.

More generally, since the desired distribution from which to sample states can vary according to the situation at hand, one needs a flexible approach to sampling from the state space. Here, we discuss general methods, adapted from statistics literature to quantum problems,[6] that can be applied to arbitrary distributions. We investigate two sampling strategies: independence sampling [8] and the Markov-chain Monte Carlo (MCMC) method [8–10]. In independence sampling, one generates sample points independently and randomly according to some convenient distribution, and then uses either rejection sampling or importance sampling for approaching the target sample. This algorithm is very simple and straightforward to implement, but can become inefficient for problems where the convenient distribution from which samples are generated is too far from the target distribution. The problems of independence sampling are remedied by the MCMC method, where sample points are generated by means of a Markov-chain random walk, making use of the current sample point to decide on a clever choice for the next sample point so that one approaches the target sample efficiently.

Both sampling strategies—independence sampling and the MCMC algorithm—are useful for all situations that one may encounter, i.e., any dimension, any choice of measurement, even if one does not possess an explicit parameterization of the domain of integration (i.e., the reconstruction space; see section 2). If one has a parameterization of the integration domain, one can also make use of Hamiltonian Monte Carlo (HMC) methods to more efficiently generate a good sample; HMC methods are discussed in [11], and our companion paper [12] deals with applying them to sampling in the reconstruction space.

Samples generated by independence sampling or by MCMC are applied in various examples for illustration. We find the marginal distribution for the purity of qubit, qutrit, or qubit-pair states (figures 2(b) and 3), determine the probability that a two-qubit state of known purity is separable (figure 2(a)), and find the sizes of regions with bounded likelihood for qubit-pair data from a tomographically incomplete measurement (figure 5).

## 2. Priors and constraints

A general measurement in quantum mechanics is a probability-operator measurement (POM).[7] The POM has outcomes $\Pi_1, \Pi_2, \ldots, \Pi_K$, which are nonnegative operators, $\Pi_k \geqslant 0$, with unit sum, $\sum_{k=1}^{K} \Pi_k = 1$. If $\rho$ is the true state, the probability that the $k$th detector clicks is given by the Born rule,

$$p_k = \mathrm{tr}\left\{\Pi_k \rho\right\}. \tag{1}$$

All the possible probabilities $p = (p_1, p_2, \ldots, p_K)$ for the chosen POM constitute the *probability space*. Given $p$, it is customary to report a $\rho$ for which (1) holds. If there is a choice among several $\rho$s for the same $p$ (which can happen when the POM is not informationally complete), we pick one representative [13], and so have a one-to-one mapping $\rho \leftrightarrow p$. These $\rho$s constitute the *reconstruction space* $\mathcal{R}_0$. Because of the one-to-one mapping between states and probabilities, we will identify $p$ with $\rho$, and regions in the probability space with corresponding regions in the reconstruction space. Note that, while the probability space is always convex, it may not be possible to choose a convex reconstruction space.[8]

The measurement data $D$ consist of the observed sequence of detector clicks, with $n_k$ clicks of the $k$th detector after measuring a total number of $N = \sum_{k=1}^{K} n_k$ copies of the state. The probability of obtaining $D$, if $p$ is the true state, is the point likelihood

$$L(D|p) = p_1^{n_1} p_2^{n_2} p_3^{n_3} \cdots p_K^{n_K}. \tag{2}$$

$L(D|p)$ takes on its largest value for the maximum-likelihood estimator [14],

$$L_{\max}(D) \equiv \max_{p \in \mathcal{R}_0} L(D|p) \equiv L\left(D\Big|\widehat{p}_{\mathrm{ML}}\right). \tag{3}$$

The positivity of $\rho$ and its normalization to unit trace ensure that $p$ satisfies the basic constraints $p_k \geqslant 0$ and $\sum_k p_k = 1$. Since the probabilities result from the POM via the Born rule, the positivity of $\rho$ usually implies further constraints on $p$. For example, consider a qubit measured by a three-outcome trine POM with

---

[6] Ironically, the statisticians had earlier learned the methods from physicists.

[7] In literature, POM is often written as POVM, standing for 'positive operator-valued measure'. This reference to the mathematical discipline of measure theory arises for historical reasons. We prefer the physical term POM as a more descriptive and relevant here.

[8] For example, if the probabilities determine all nondiagonal elements of the $3 \times 3$ matrix of a qutrit state, it is not possible to assign diagonal matrix elements such that the reconstruction space is convex.

probabilities

$$p_1 = \frac{1}{3}(1+x), \qquad \left.\begin{array}{c} p_2 \\ p_3 \end{array}\right\} = \frac{1}{3}\left(1 - \frac{1}{2}x \pm \frac{\sqrt{3}}{2}y\right), \tag{4}$$

where $x = \langle\sigma_x\rangle$ and $y = \langle\sigma_y\rangle$ are expectation values of Pauli operators. These trine probabilities are further constrained by $\sum_{k=1}^{3} p_k^2 \leq \frac{1}{2}$. Quite generally, there are such additional constraints for the probabilities, and it may not be easy or feasible to state them explicitly for high-dimensional systems measured by many-outcome POMs. A $p$ is called *physical* or *permissible* if it satisfies all these constraints.

We summarize all constraints in $w_{\text{cstr}}(p)$, which is a product of step functions and delta functions, and vanishes if $p$ is not permissible. For example, for the basic constraints, we have[9]

$$w_{\text{basic}}(p) \equiv \eta\left(p_1\right)\eta\left(p_2\right)\cdots\eta\left(p_K\right)\delta\left(\sum_k p_k - 1\right) \tag{5}$$

as a factor in

$$w_{\text{cstr}}(p) = w_{\text{basic}}(p)\, w_{\text{qu}}(p), \tag{6}$$

where $w_{\text{qu}}(p)$ is the product of step functions that specify the constraints imposed by quantum mechanics. Then, the volume element $(\mathrm{d}p)$ of the probability space is

$$(\mathrm{d}p) = \mathrm{d}p_1 \mathrm{d}p_2 \ldots \mathrm{d}p_K \; w_{\text{cstr}}(p), \tag{7}$$

and we have

$$(\mathrm{d}\rho) = (\mathrm{d}p)\, w_0(p) \tag{8}$$

for the volume element of the infinitesimal vicinity of state $\rho$ in $\mathcal{R}_0$, where $w_0(p)$ is the (unnormalized) prior density of our choice. Specifically, we will discuss two choices for $w_0(p)$ in the examples later. The first is the *primitive prior*,

$$w_{\text{primitive}}(p) = 1, \tag{9}$$

so that the density is uniform in $p$ over the (physical) probability space. The second is known as the *Jeffreys prior* [15],

$$w_{\text{Jeffreys}}(p) = \frac{1}{\sqrt{p_1 p_2 \cdots p_K}}, \tag{10}$$

which is a common choice of prior when no external prior information is available [1, 2].

Upon multiplying the prior density with the point likelihood for the observed data, we get the (unnormalized) posterior density

$$w_D(p) = w_0(p) L(D|p). \tag{11}$$

While sampling from the probability space in accordance with the prior $w_0(p)$ is often relatively easy, sampling according to the posterior $w_D(p)$ is usually difficult. With the prior and posterior densities at hand, we can now define the size and credibility of a region $\mathcal{R}$ in the reconstruction space. The computation of these region-specific values is an important application of random samples in the context of quantum state estimation [16]; an example is given in section 6.

The *size* $S_{\mathcal{R}}$ of region $\mathcal{R}$ is

$$S_{\mathcal{R}} = \frac{\int_{\mathcal{R}}(\mathrm{d}\rho)}{\int_{\mathcal{R}_0}(\mathrm{d}\rho)} = \frac{\int_{\mathcal{R}}(\mathrm{d}p)\, w_0(p)}{\int_{\mathcal{R}_0}(\mathrm{d}p)\, w_0(p)}, \tag{12}$$

and

$$C_{\mathcal{R}} = \frac{\int_{\mathcal{R}}(\mathrm{d}p)\, w_D(p)}{\int_{\mathcal{R}_0}(\mathrm{d}p)\, w_D(p)} \tag{13}$$

is its *credibility*. $S_{\mathcal{R}}$ is the prior content of region $\mathcal{R}$, i.e., the probability that the true state is in $\mathcal{R}$ before any data are acquired; $C_{\mathcal{R}}$ is its posterior content, i.e., the probability that the true state is in $\mathcal{R}$ conditioned on the data.[10]

---

[9] There are POMs for which partial sums of the $p_k$s have fixed values, in which case there is more than one delta function in $w_{\text{basic}}(p)$; see, e.g., (23) in [13] or (30) and (36) in [12]. One identifies these situations easily and we need not elaborate on them.

[10] The conventions used here differ somewhat from those in [13]. In particular, we include here the factor $w_{\text{cstr}}(p)$ in $(\mathrm{d}p)$ and we prefer unnormalized prior and posterior densities in the current context, so that the denominators in (12) and (13) do not have unit values.

If we wish to sample the reconstruction space in accordance with the prior $w_0(p)$, the fraction of sample points in region $\mathcal{R}$ should be equal to the size $S_{\mathcal{R}}$ of the region; and equal to the credibility $C_{\mathcal{R}}$ when sampling according to the posterior $w_D(p)$. The situation can be reversed: we may have a procedure for generating a random sample, and then this procedure defines the underlying prior; an example is 'prior I' in section 4.

In the context of quantum state estimation, one needs to compute $S_{\mathcal{R}}$ and $C_{\mathcal{R}}$ for given region $\mathcal{R}$ and data $D$. Since the quantum constraints are generally highly nontrivial, leading to a probability space with complicated boundaries, and the dimension of the probability space grows rapidly (as the square of the dimension for tomographically complete measurements), the integrals in $S_{\mathcal{R}}$ and $C_{\mathcal{R}}$ are difficult to compute directly. The structure of the integrals naturally suggests the use of Monte Carlo methods for evaluation: generate points with a density distributed according to a target $w(p)$ (in our case, $w_0(p)$ or $w_D(p)$); the size and credibility are then the ratio of the number of points contained in $\mathcal{R}$ to the total number of points in the full reconstruction space $\mathcal{R}_0$.

One also needs a systematic and efficient way of numerically checking if a given $p$ satisfies all quantum constraints, i.e., if the given $p$ is physical—this is required as part of the procedure for evaluating the constraint factor $w_{\mathrm{cstr}}(p)$. For a high-dimensional system measured by a many-outcome POM, the constraints cannot easily be expressed explicitly in terms of inequalities, and can thus only be checked numerically. If the POM is informationally complete, the $p \leftrightarrow \rho$ mapping is linear and usually known quite explicitly and one could just check if it gives a nonnegative $\rho$. For other POMs, this approach is often not available because the $p \leftrightarrow \rho$ mapping is involved and only defined for physical $p$s, and then other methods must be used.

In appendix A, we provide an algorithm for checking the physicality of a given $p$. We make use of a figure-of-merit functional $Q(p; \hat{p})$, where $p$ is the probability (assumed to satisfy the basic constraints) to be checked for physicality and $\hat{p}$ is a random variable in the reconstruction space, i.e., $\hat{p}$ is physical. $Q$ is chosen such that it attains its optimal value when $\hat{p}$ is as close to $p$ as possible. Given $p$, one optimizes $Q$ over $\hat{p}$ using gradient methods, and if the optimal $\hat{p}$ is equal to $p$, $p$ is physical; otherwise, it is not.

## 3. Independence sampling: rejection sampling and importance sampling

In independence sampling, as the name suggests, sample points are randomly generated independently of one another. In general, it is not straightforward to sample directly from the target distribution $w_t(p)$—here equal to $w_{\mathrm{cstr}}(p)w_0(p)$ or $w_{\mathrm{cstr}}(p)w_D(p)$ if sampling in accordance with the prior or the posterior. However, as long as we can sample over the probability space (perhaps using a convenient parameterization) with a known reference distribution $w_r(p)$, we can approach the target distribution by means of rejection sampling or importance sampling [8]. The factor $r(p)$ that relates the target distribution to the reference distribution,

$$w_t(p) = w_r(p)\, r(p), \tag{14}$$

can be regarded as the ratio '$r(p) = w_t(p)/w_r(p)$,' but this should be done with care since $w_t(p)$ and $w_r(p)$ usually share singularities of the delta-function kind.

The easiest way of sampling the probability space is often to first sample uniformly in $p$ from the space of probabilities that satisfy only the basic constraints of positivity and unit sum, as specified by the factor $w_{\mathrm{basic}}(p)$ of (5). We refer to this space as the basic probability simplex, and the (unnormalized) sampling distribution $w_r(p)$ is equal to $w_{\mathrm{basic}}(p)$ from (5), i.e., it takes a constant value over the entire simplex, and is zero for $p$s violating the basic constraints. In appendix B, we provide two algorithms for sampling from this $w_r(p)$. The physical probability space is a subregion of this simplex, with the additional quantum constraints imposed by the rejection or importance sampling procedures.

In *rejection sampling*, we draw many sample points according to the chosen reference distribution $w_r(p)$, and then reject (i.e., discard) or accept points in such a way that the remaining sample points are distributed according to the target distribution $w_t(p)$. More specifically, we accept a sample point $p^{(j)}$ with probability

$$a = \frac{r\left(p^{(j)}\right)}{R}, \tag{15}$$

where $R \equiv \max_p \{r(p)\}$. One calls $a$ the *acceptance ratio*.

Rejection sampling requires one to discard points in accordance with the acceptance ratio, and one ends up with fewer sample points than the initial set drawn from $w_r(p)$. In *importance sampling*, instead of discarding points, one attaches a weight to each point to compensate for the difference between the sampling and the target distributions. For sample point $p^{(j)}$, the weight is

$$W_j = r\left(p^{(j)}\right). \tag{16}$$

This weight can be thought of as a multiplicity for each sample point in accordance with the target $w_t(p)$, so that each point $p^{(j)}$ counts $W_j$ times in computing the ratio of number of points in $\mathcal{R}$ to the total number of points in $\mathcal{R}_0$ for the value of the integral $S_\mathcal{R}$ or $C_\mathcal{R}$. Moreover, the weights should have finite variance for good practical performance and ideally be bounded [17]. But this can be hard to check.

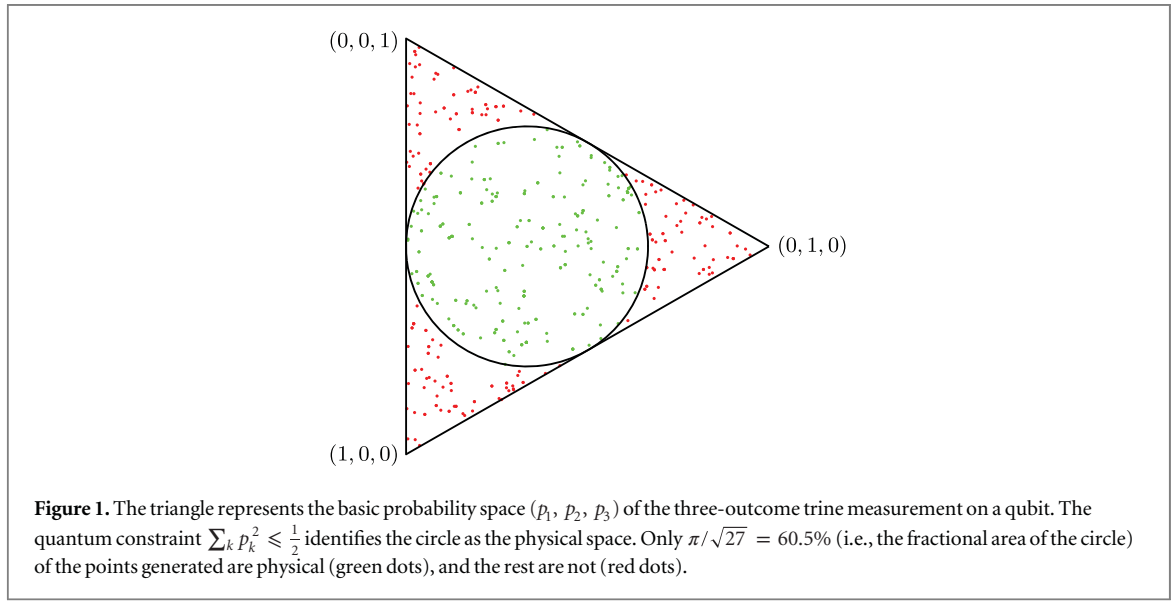

**Figure 1.** The triangle represents the basic probability space $(p_1, p_2, p_3)$ of the three-outcome trine measurement on a qubit. The quantum constraint $\sum_k p_k^2 \leq \frac{1}{2}$ identifies the circle as the physical space. Only $\pi/\sqrt{27} = 60.5\%$ (i.e., the fractional area of the circle) of the points generated are physical (green dots), and the rest are not (red dots).

For the reference distribution that is uniform on the simplex, $w_r(p) = w_{\text{basic}}(p)$, we have $r(p) = w_{\text{qu}}(p) w(p)$ in (15) and (16) with $w(p) = w_0(p)$ for prior sampling or $w(p) = w_D(p)$ for posterior sampling. Both in rejection sampling and in importance sampling, unphysical points, i.e., those that satisfy the basic but not the quantum constraints, do not contribute to the integral, since they are either rejected with unit probability in rejection sampling, or carry zero weight in importance sampling. This means that, if $\mathcal{R}_0$ is a small subregion of the basic probability simplex, one ends up with only a small fraction of the sample points contributing finally to the integral. For example, for a three-outcome trine measurement on the single qubit of (4), only $\pi/\sqrt{27} = 60.5\%$ of the points sampled from $w_r(p) = w_{\text{basic}}(p)$ are physical (see figure 1). The yield decreases as the dimensionality of the system increases: for a nine-outcome trine-antitrine (TAT) measurement on a qubit-pair (see section 6), only about 10% of the points are physical [13]; and for the sixteen-outcome POM used in section 4, only one in 50 000 candidate points is accepted.

One also runs into problems where the ratio $r(p)$ is sharply peaked. For example, the Jeffreys prior formally becomes infinite when (at least) one of the $p_k$s is zero. In practice, one never gets a sample point $p^{(j)}$ with a $p_k$ that is *exactly* zero, so that $r(p^{(j)})$ is never infinite. Still, any sample point in the vicinity of the singular points will have a very large $r(p^{(j)})$ value. The normalization constant $R$ for the Jeffreys prior is also formally infinite, calling to question the applicability of the rejection sampling procedure. In practice, one can take $R$ as a large constant by approximating the target Jeffreys prior by one with a 'cutoff' value when one or more of the $p_k$s vanish. This still, however, makes the acceptance rate tiny for all $p$s away from the singular points. Correspondingly, in importance sampling, large weights are attached to the points in the vicinity of these singular points, and the main contribution to the integral then comes from just those few points.

Both the problems of small physical subregion and sharply peaked priors stem from the fact that the target distribution can be very different from the reference distribution. Whenever possible, one should start with samples from a $w_r(p)$ that is close to $w_t(p)$. Nevertheless, independence sampling according to a uniform $w_r(p)$ on the basic probability simplex is straightforward to set up, and can provide an easy first estimate of the desired integral, or more generally, a rough first sample.

## 4. Example: volume of separable two-qubit states

For a first application, we sample the two-qubit state space and ask how large the volume of the set of separable states (or conversely, entangled states) is. In [3], a natural prior on the set of states is used that is induced by the Haar measure on the group of unitary matrices and the Lebesgue measure on the real space (labeled 'Prior I' in figures 2 and 3). For this prior, numerical results establish that $63.2\% \pm 0.2\%$ of the mixed two-qubit states are separable.

Here, we consider the scenario where each of the two qubits is measured by the four-outcome tetrahedron POM of [18] separately. The resulting two-qubit POM (which is informationally complete) has sixteen outcomes with the single constraint of unit sum, so the probability space is fifteen-dimensional. We employ the simple rejection sampling method to generate probabilities in accordance with the primitive prior (labeled 'Prior II' in figures 2 and 3). Altogether 53 332 physical probabilities are generated, with an acceptance rate of 0.00215%.[11] This small acceptance rate is due to the tiny fraction of the physical region compared to the whole probability simplex, which is fifteen-dimensional. Then we construct the corresponding density matrices as well as their partial transposes. The well-known Peres–Horodecki criterion states that if a state $\rho$ is separable, then its partial transpose has nonnegative eigenvalues; otherwise $\rho$ is entangled. According to our numerical results, the probability that a randomly generated two-qubit state is separable equals 24.2% ± 0.2%, which is much smaller than the value reported in [3]. Clearly, the two priors are quite different.

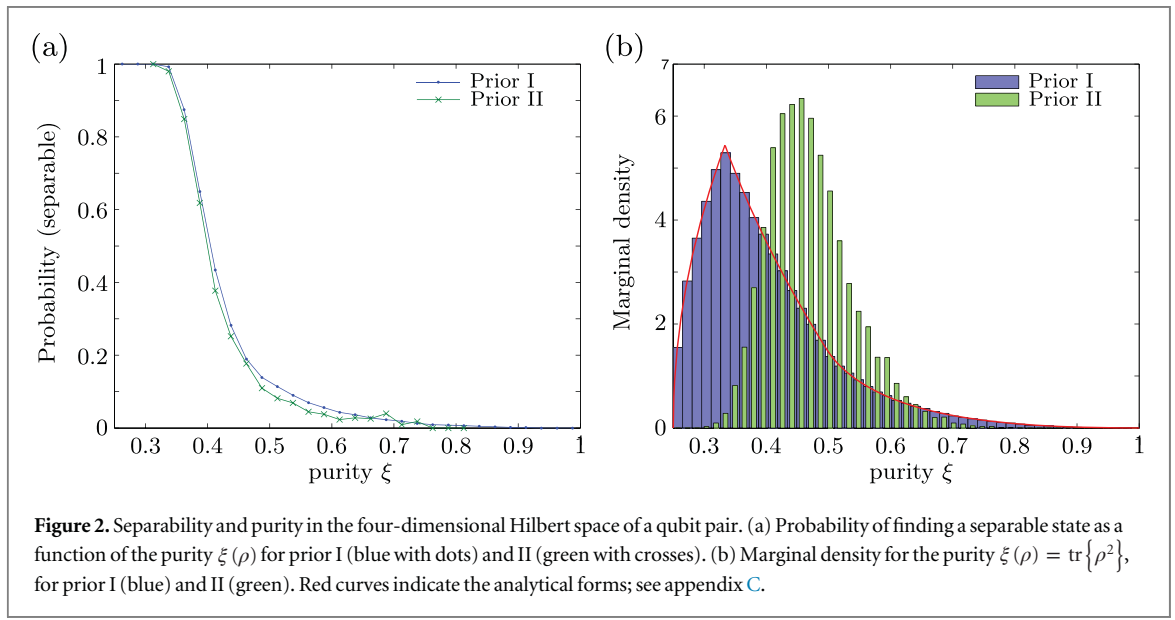

**Figure 2.** Separability and purity in the four-dimensional Hilbert space of a qubit pair. (a) Probability of finding a separable state as a function of the purity $\xi(\rho)$ for prior I (blue with dots) and II (green with crosses). (b) Marginal density for the purity $\xi(\rho) = \mathrm{tr}\left\{\rho^2\right\}$, for prior I (blue) and II (green). Red curves indicate the analytical forms; see appendix C.

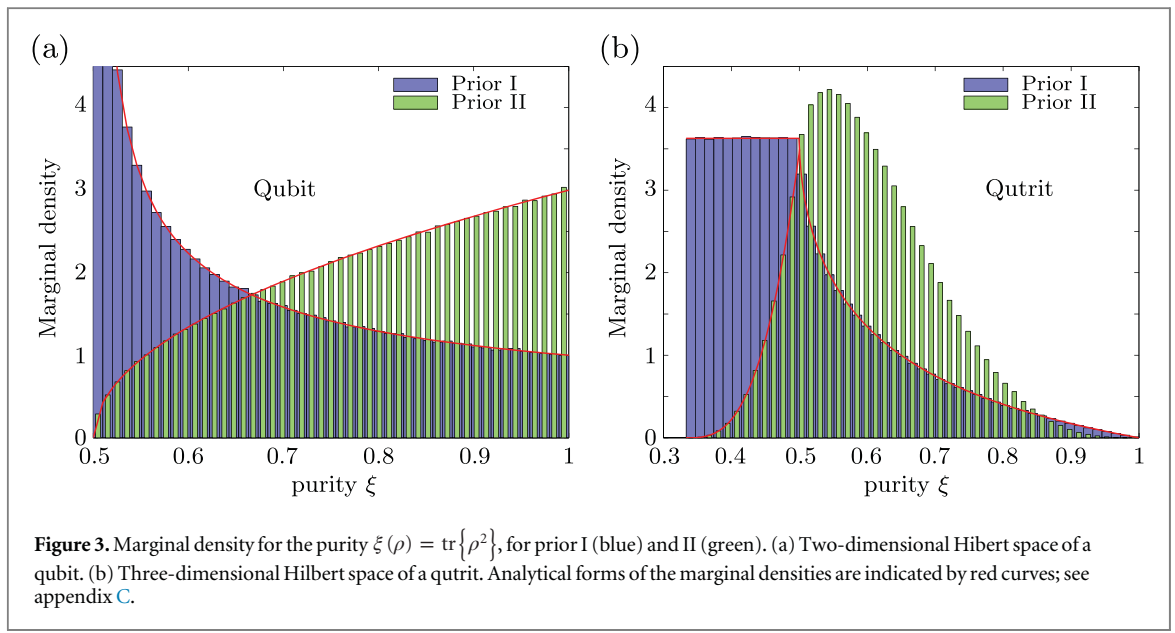

**Figure 3.** Marginal density for the purity $\xi(\rho) = \mathrm{tr}\left\{\rho^2\right\}$, for prior I (blue) and II (green). (a) Two-dimensional Hibert space of a qubit. (b) Three-dimensional Hilbert space of a qutrit. Analytical forms of the marginal densities are indicated by red curves; see appendix C.

To better understand how these priors differ, and how this difference affects the computation of the volume of separable states, it is worth considering the physical connection between the purity $\xi(\rho) = \mathrm{tr}\left\{\rho^2\right\}$ of the states and their separability. In figure 2(a), we show the probability of finding a separable two-qubit state as a function of the purity for both priors. Although the two curves are very similar, there appears to be a systematic difference between the two priors: for given purity, states are a bit more likely to be separable for prior I than prior II. The strong similarity, however, tells us that the prior densities conditioned on the purity are close. But the marginal densities for the purity are rather different; see figure 2(b). For our prior II, we find the marginal density peaking at a higher purity value, indicating that this prior puts more weight on the states of higher purity and less weight on the low purity states. This, together with the fact that higher purity states are less likely to be separable, results in a smaller overall probability for our prior to produce a separable state.

---

[11] The same random sample according to the primitive prior can also be obtained as follows [5]: generate a square random matrix $A$ with all entries being independent complex Gaussian numbers, and compute $\rho = AA^{\dagger}/\mathrm{tr}\left\{AA^{\dagger}\right\}$, which is automatically physical. This procedure is much faster, but only applies to informationally complete POMs with the primitive prior.

To further see the difference between these two approaches, we also plot the probability density of the quantum states for qubit as well as qutrit state space in figure 3. Analytical forms of the marginal densities are indicated in the plots by red curves; see appendix C for details.

## 5. MCMC sampling

The problems of independence sampling—the low yield in high-dimensional spaces and the difficulty in handling sharply peaked distributions reliably—can be resolved by using the MCMC strategy (see, for instance, [19]). In MCMC, sample points that obey the basic constraints are generated sequentially, with the position of the next point depending on the position of the current point; hence the term 'Markov chain.' One makes use of a random walk such that the next sample point is likely to be in the vicinity of the current point; this gives a high chance of staying within the permissible region if the current point is physical, or within the same peak if the current point is within a sharply peaked region of

$$\tilde{w}(p) = w_{qu}(p) w(p), \tag{17}$$

with $w(p) = w_0(p)$ or $w(p) = w_D(p)$.

The Markov chain's stationary distribution is to be the target distribution. To achieve this, the Metropolis–Hastings Monte Carlo (MHMC) procedure [9, 10] is adopted when performing the random walk; see appendix D for a review. For our purposes, it is expedient to reparameterize the probability space: let $x = \{x_1, x_2, \cdots, x_K\}$ be such that $p_k = x_k^2$ for all $k$. Since the $p_k$s add up to 1, we have $\sum_{k=1}^{K} x_k^2 = 1$, so that the space of $x$ is the surface of the unit $(K-1)$-hypersphere centered about the origin. From point $x$, we then propose a new point by drawing a $K$-dimensional multivariate Gaussian random variable with mean $x$ and variance $\sigma^2$, and normalize it back to unit length. The symmetry of such a proposal distribution is guaranteed by the spherical symmetry of the Gaussian distribution.

The reparameterization in terms of $x$ requires a corresponding transformation of the target distribution,

$$(\mathrm{d}p)\,\tilde{w}(p) = (\mathrm{d}x)\,\tilde{w}(p)\left|\frac{\mathrm{d}p}{\mathrm{d}x}\right|_{p=x^2} \propto (\mathrm{d}x)\,|x|\,\tilde{w}\big(p = x^2\big). \tag{18}$$

Also, the starting point of the random walk should be a physical probability, which can be ensured by picking a state $\rho$ (the maximally mixed state, for instance), computing the corresponding probabilities $p^{(1)}$, and setting the initial $x^{(1)} = \sqrt{p^{(1)}}$.

Putting it all together, the $x$-parameterized MHMC scheme is as follows:

**xMHMC1** Pick an arbitrary state. Obtain $p^{(1)}$ from the Born rule, then set $x^{(1)} = \sqrt{p^{(1)}}$ and $j=1$.

**xMHMC2** Randomly generate $\Delta x$, a $K$-dimensional variable with mean 0 and variance $\sigma^2$.

**xMHMC3** Compute $x^\star = \frac{x^{(j)} + \Delta x}{|x^{(j)} + \Delta x|}$ and $p^\star = \left(x^\star\right)^2$.

**xMHMC4** Compute the acceptance ratio

$$a = \min\left\{\frac{\sqrt{p^\star}\,\tilde{w}\big(p^\star\big)}{\sqrt{p^{(j)}}\,\tilde{w}\big(p^{(j)}\big)},\, 1\right\}. \tag{19}$$

**xMHMC5** Draw a random number $b$ uniformly from the range $0 < b < 1$. If $a > b$, set $p^{(j+1)} = p^\star$; otherwise, set $p^{(j+1)} = p^{(j)}$.

**xMHMC6** Update $j \to j + 1$. For target number of sample points $M$, escape the loop if $j=M$; otherwise, return to xMHMC3.

Some attention should be paid to the choice of variance $\sigma^2$ in xMHMC2. The corresponding standard deviation $\sigma$ can be viewed as the typical 'step size' in the random walk. Generally, if the step size is too large, the acceptance rate tends to be low, since a single step may take one too far from the permissible or important region; if the step size is too small, the random walk takes a long time to explore the entire space. Therefore, $\sigma$ has to be chosen carefully. In statistics literature, using general arguments invoking the central limit theorem in the many-parameter situation, one is told that the optimal $\sigma$ should be chosen such that acceptance rate is around 23% (see, for instance, [20] and [21]). This gives a good rule of thumb for choosing the step size $\sigma$, and turns out to be quite reasonable even for small-dimensional problems; see figure 4.

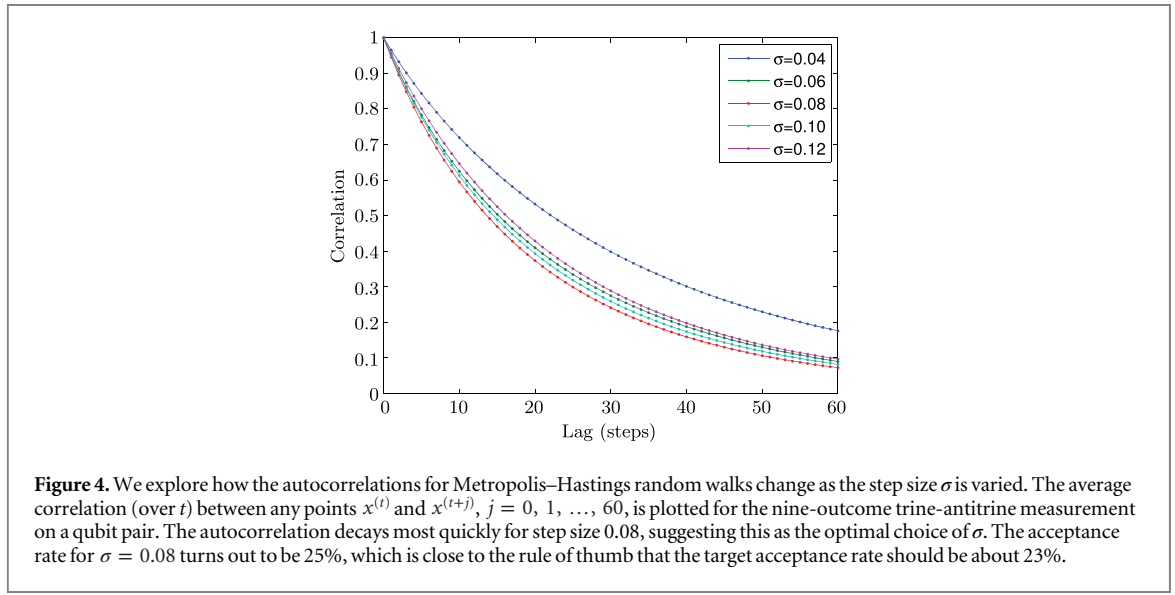

**Figure 4.** We explore how the autocorrelations for Metropolis–Hastings random walks change as the step size $\sigma$ is varied. The average correlation (over $t$) between any points $x^{(t)}$ and $x^{(t+j)}$, $j = 0, 1, \ldots, 60$, is plotted for the nine-outcome trine-antitrine measurement on a qubit pair. The autocorrelation decays most quickly for step size 0.08, suggesting this as the optimal choice of $\sigma$. The acceptance rate for $\sigma = 0.08$ turns out to be 25%, which is close to the rule of thumb that the target acceptance rate should be about 23%.

**Table 1.** CPU time taken to generate the sample of 100 000 (physical) probabilities with different sampling strategies for qubit pairs measured by the TAT POM in accordance with the primitive prior. DG: direct gradient; CG: conjugate gradient; see appendix A.

| Sampling scheme | CPU time |
| --- | --- |
| Independence with DG | 14 hr |
| MCMC with DG | 1 hr 20 min |
| MCMC with CG | 11 min |
| MCMC with parameter searching | 13 min |

## 6. Example: incomplete two-qubit tomography

For a comparison of the differences between various sampling methods, we consider the TAT scheme of [22] (see also section 6 in [13]) for quantum key distribution. This TAT scheme can be implemented as follows: begin with a source of entangled qubit pairs and distribute one qubit each to the two communicating particles; one qubit is measured by the trine POM of (4); the other is measured by the antitrine POM, which is (4) with the signs of $x$ and $y$ reversed. The two-qubit POM has nine outcomes subject to the single constraint of unit sum, resulting in an eight-dimensional probability space. For the simulated experiments, we measure 60 copies of qubit pairs, and the data, in one experiment, are $D = \{11, 4, 5, 2, 10, 5, 4, 6, 13\}$, which are used in the specification of the optimal error regions below.

We generated various sets of samples in accordance with the primitive prior as well as with the Jeffreys prior. The platform used for generating these samples is a standard desktop (Intel i7–3770 CPU, with quad core and 8 GB RAM). The CPU time taken to generate the sample of 100 000 (physical) probabilities with different sampling strategies is summarized in table 1. We also show the time taken by MCMC, but with a physicality check that exploits the structure of the TAT (labeled 'MCMC with parameter searching' in table 1). In the TAT version of the matrix in (47) of [12], eight out of nine real parameters for a reconstruction space can be determined by the probabilities, with the last one being in the range $[-1, 1]$. If such a ninth parameter can be found to make the corresponding state positive semidefinite, then the generated $p$ is physical; otherwise it is not. The CPU time by this procedure is almost the same as that by MCMC with CG in the physicality check. Moreover, there is barely any difference in time whether the sample is generated according to the primitive prior or the Jeffreys prior, as the time-consuming part is the checking of physicality of the generated probabilities.

In figure 5, we show the size $s_\lambda$ as a function of $\lambda$ for different regions for data $D$ using samples of various sizes. Here, $\lambda$ is the likelihood threshold for the region, with $0 \leq \lambda \leq 1$. The region specified by a given $\lambda$ is the set of points with point likelihoods satisfying $L(D|p) \geq \lambda L(D|\hat{p}_{ML})$. For figure 5(a) using the samples generated by independence sampling, there is not much difference for the curves obtained with the samples of sizes 10 000 and 100 000 for both the primitive prior and the Jeffreys prior. However, for figure 5(b) using the samples generated by MCMC, the curve obtained with 10 000 sample points deviates quite far from the curves obtained with larger sample sizes for the primitive prior. A possible reason is that the chain may have been 'trapped' at the mode of the prior in the transformed $x$ space for a significant portion of the time, and the sample with 10 000 points is not large enough for the random walk to reach the whole space. This does not happen for the Jeffreys prior, which is flat in the $x$ space.

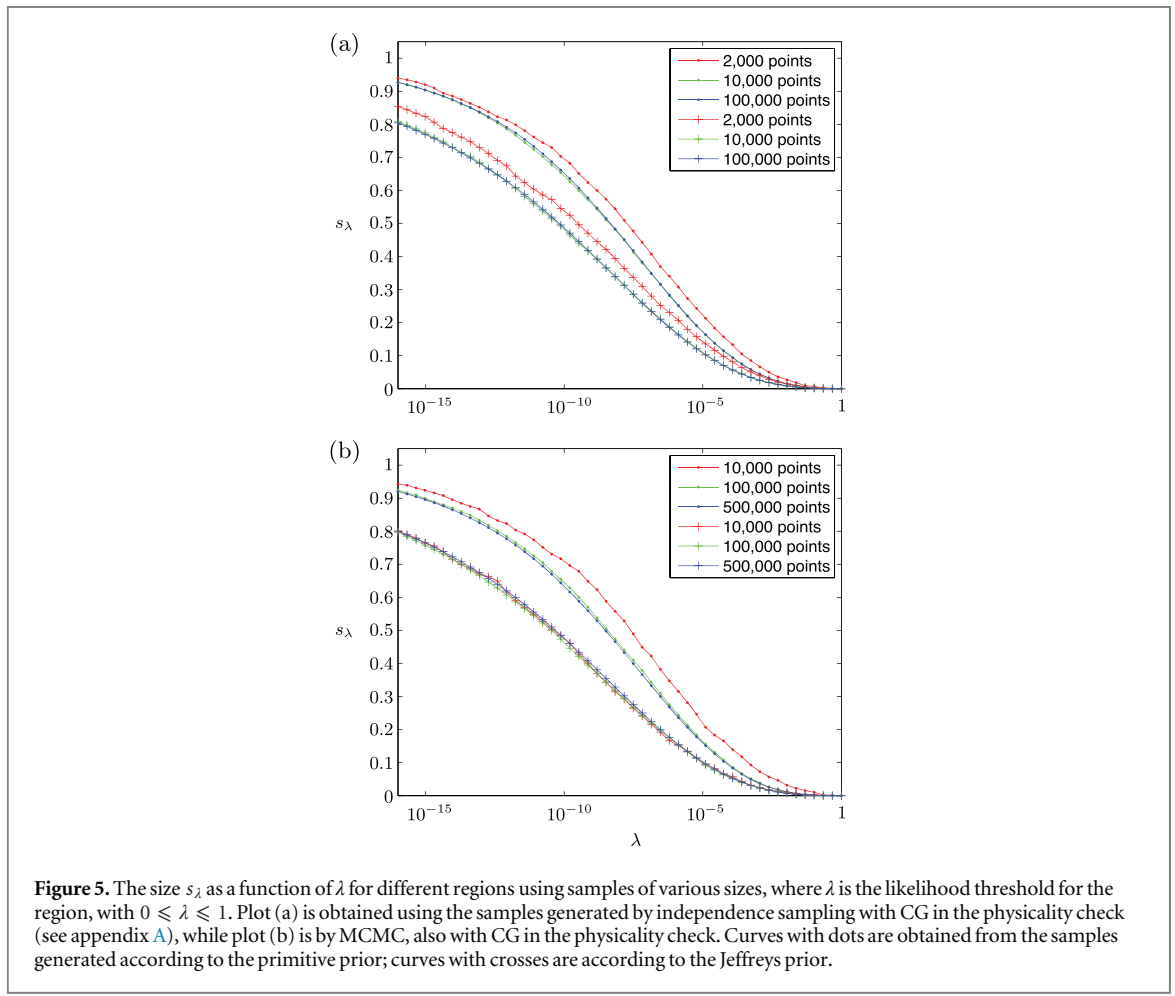

**Figure 5.** The size $s_\lambda$ as a function of $\lambda$ for different regions using samples of various sizes, where $\lambda$ is the likelihood threshold for the region, with $0 \leqslant \lambda \leqslant 1$. Plot (a) is obtained using the samples generated by independence sampling with CG in the physicality check (see appendix A), while plot (b) is by MCMC, also with CG in the physicality check. Curves with dots are obtained from the samples generated according to the primitive prior; curves with crosses are according to the Jeffreys prior.

In terms of the 'quality' of the sample points, we notice that a sample of size 10 000 generated in accordance with the primitive prior by independence sampling is good enough to produce figure 5; while it requires a sample of size 100 000 if we use MCMC. In this sense, the comparison of the time taken to generate the same number of physical probabilities in table 1 may not always be fair. At least for our particular purpose of calculating the sizes, both the independence sampling and the MCMC are roughly equivalent. The lower acceptance rate of independence sampling is supplemented by a faster convergence rate, while MCMC converges more slowly at the benefit of a higher acceptance rate.

## 7. Conclusion

We have shown how one can perform rejection sampling, importance sampling, and MCMC sampling in the probability space (and thus also in the reconstruction space) with due attention paid to all the constraints obeyed by physical probabilities. Rejection sampling and importance sampling are rather simple to implement but they have a low yield and are costly (in CPU time) unless one manages to check the physicality of candidate probabilities in an efficient way. While MCMC sampling tends to be less costly because the yield is higher (fewer candidate probabilities rejected), this comes at the price of correlations in the sample, which in turn requires larger samples to achieve the same accuracy that rejection sampling and importance sampling get for smaller samples. For comparison, we have generated samples of various sizes in the probability space for two-qubit states measured by an incomplete POM. Using the samples created, the sizes for different regions are then calculated.

Once the samples are at hand, one can now efficiently compute the optimal error regions for quantum state estimation introduced in [13], where integrals over high-dimensional regions in the quantum state space must be evaluated. While this application motivated these investigations, the random samples themselves can be used for the many purposes mentioned in the introduction. The algorithms explored here—independence sampling and MCMC sampling—can be applied to problems of any dimension, any choice of measurement, and any target distribution, even if one does not have an explicit parameterization of the state space, as is often the case. This flexibility is of great practical value.

If an explicit parameterization of the state space is available, one can alternatively sample by the so-called HMC algorithm [11]. HMC is very different from both independence sampling and MCMC in many aspects. We deal with HMC in our companion paper [12].

## Acknowledgments

We thank Michael Evans and Yong Siah Teo for stimulating discussions. This work is funded by the Singapore Ministry of Education (partly through the Academic Research Fund Tier 3 MOE2012-T3–1-009) and the National Research Foundation of Singapore.

## Appendix A. Determining the physicality of $p$

Inspired by maximum-likelihood considerations, we choose a figure-of-merit functional $Q(p; \hat{p})$, where $p$ is the candidate probability to be checked and $\hat{p}$ is a physical probability, such that $Q(p; \hat{p})$ has a global maximum in $\hat{p}$ but no local maxima. The idea is that $p$ is physical if $\max_{\hat{p}}\{Q(p; \hat{p})\}$ is equal to its largest possible value $\max_{p,\hat{p}}\{Q(p; \hat{p})\}$; otherwise $p$ is not physical. Examples are $Q(p; \hat{p}) = \sum_k p_k \log(\hat{p}_k/p_k) \leqslant 0$ or $Q(p; \hat{p}) = \sum_k \sqrt{p_k \hat{p}_k} - 1 \leqslant 0$, where the equal signs hold only if $\hat{p} = p$, and $\max_{\hat{p}}\{Q(p; \hat{p})\} < 0$ implies that $p$ is not physical.

The physicality of $\hat{p}$ is ensured by the Born rule $\hat{p}_k = \mathrm{tr}\{\Pi_k \hat{\rho}\} = \langle \Pi_k \rangle$, where

$$\hat{\rho} = \frac{A^{\dagger}A}{\mathrm{tr}\{A^{\dagger}A\}} \quad \text{with arbitrary } A \tag{A.1}$$

is assuredly nonnegative and has unit trace. (Whether these $\hat{\rho}$s are in the reconstruction space or not is irrelevant here.) Infinitesimal variations of $\hat{p}$ originate in variations of $\hat{\rho}$ which in turn result from varying $A^{\dagger}$ and $A$. The response of $Q(p; \hat{p})$ to such infinitesimal changes is

$$\delta Q = \frac{1}{\mathrm{tr}\{A^{\dagger}A\}} \mathrm{tr}\{\delta A^{\dagger} G + G^{\dagger} \delta A\}, \tag{A.2}$$

where

$$G = A(R - \langle R \rangle) \quad \text{with } R = \sum_k \frac{\partial Q}{\partial \hat{p}_k} \Pi_k \tag{A.3}$$

and its adjoint $G^{\dagger}$ are the components of the gradient. As in [23], steepest ascent ('follow the gradient uphill') is achieved by putting $\delta A^{\dagger} = \epsilon G^{\dagger}$ and $\delta A = \epsilon G$ with $\epsilon > 0$, when

$$\delta Q = \frac{2\epsilon \, \mathrm{tr}\{G^{\dagger}G\}}{\mathrm{tr}\{A^{\dagger}A\}} = 2\epsilon \left\langle (R - \langle R \rangle)^2 \right\rangle \geqslant 0. \tag{A.4}$$

The equal sign holds only when the gradient vanishes, $A^{\dagger}G = G^{\dagger}A = 0$ or

$$R\hat{\rho} = \hat{\rho}R = \langle R \rangle \hat{\rho}; \tag{A.5}$$

here and elsewhere, the $\hat{p}_k$s in $R$ and other expectation values $\langle \ \rangle$ always refer to the $\hat{\rho}$ under consideration.

Thus, we update $\hat{\rho}$ in accordance with

$$\hat{\rho} \to \hat{\rho}' = \frac{[1 + \epsilon(R - \langle R \rangle)]\hat{\rho}[1 + \epsilon(R - \langle R \rangle)]}{\left\langle [1 + \epsilon(R - \langle R \rangle)]^2 \right\rangle} \tag{A.6}$$

or, in discrete form,

$$\hat{\rho}_{j+1} = \frac{\left[1 + \epsilon\left(R_j - \langle R_j \rangle\right)\right]\hat{\rho}_j\left[1 + \epsilon\left(R_j - \langle R_j \rangle\right)\right]}{1 + \epsilon^2\left(\langle R_j^2 \rangle - \langle R_j \rangle^2\right)}. \tag{A.7}$$

Numerically, the value of $\max_{\hat{p}} Q(p; \hat{p})$ is reached when $\mathrm{tr}\left\{\left|(R_j - \langle R_j\rangle)\rho_j\right|\right\} \leqslant \varepsilon$, where

$\mathrm{tr}\{|M|\} = \mathrm{tr}\left\{\sqrt{M^\dagger M}\right\}$ is the trace norm for operator $M$ and $\varepsilon$ is a pre-chosen precision. Upon comparing this value for $\max_{\hat{p}} Q(p; \hat{p})$ with $\max_{p,\hat{p}} Q(p; \hat{p})$, we conclude that $p$ is physical or not.

This search for the maximum of $Q$ can be carried out by simply following the direct gradient (DG) $G$ of (A.2) and (A.3). Or, we can use the more efficient conjugate-gradient (CG) method [24], with the CG $H$ defined recursively,

$$H_{j+1} = G_{j+1} + \gamma_{j+1} H_j, \tag{A.8}$$

where $G_{j+1}$ is the direct gradient and $H_1 = G_1$. The parameter $\gamma_{j+1}$ is known as the Polak–Ribière criterion determined by $G$, see (A.10). Here is an outline of the iterative algorithm that employs the CG method for the physicality check (PC) of $p$, for a $d$-dimensional quantum system:

**PCCG1** Start with $j = 1$, two fixed constants $\epsilon$ and $\xi$, the $d \times d$ identity matrix $A_1 = 1$, and the maximally mixed state $\hat{\rho}_1 = 1/d$.

**PCCG2** Compute $R_1$ of (A.3) and set $G_1 = A(R_1 - \langle R_1\rangle)$, $H_1 = G_1$.

**PCCG3** Escape the loop if $\mathrm{tr}\left\{\left|(R_j - \langle R_j\rangle)\rho_j\right|\right\} \leqslant \varepsilon$; otherwise, proceed with PCCG4–8.

**PCCG4** Set $A_{j+1} = A_j + \epsilon H_j$ and compute

$$\hat{\rho}_{j+1} = \frac{A_{j+1}^\dagger A_{j+1}}{\mathrm{tr}\left\{A_{j+1}^\dagger A_{j+1}\right\}}. \tag{A.9}$$

**PCCG5** Compute $R_{j+1}$ according to (A.3) and set $G_{j+1} = A_{j+1}\left(R_{j+1} - \langle R_{j+1}\rangle\right)$.

**PCCG6** Compute

$$\gamma_{j+1} = \max\left\{\frac{\mathrm{tr}\left\{G_{j+1}^\dagger\left(G_{j+1} - \chi G_j\right)\right\}}{\mathrm{tr}\left\{G_j^\dagger G_j\right\}}, 0\right\}. \tag{A.10}$$

**PCCG7** Set $H_{j+1} = G_{j+1} + \gamma_{j+1} H_j$.

**PCCG8** Update $j \to j + 1$ and repeat the iteration from PCCG3.

Instead of using a constant $\epsilon$ in PCCG4, we employ a suitable line-optimization procedure to speed up the algorithm [24]. Such a line search can in principle expedite the optimization, but may become impractical in higher dimensions as the evaluation of many large matrices is computationally very expensive. In this case, a fixed value of $\epsilon$ is used instead. As for the parameter $\chi$ used in PCCG6, it should be chosen from the range $0 \leqslant \chi < 1$, with smaller values typically leading to quicker convergence. Compared with the direct-gradient method, each iteration of the CG method is slightly more expensive computationally. However, the number of iterations required for convergence is much smaller for the CG method. The overall effect is that the CPU time required by the CG method is a fraction of that required by the direct-gradient method.

## Appendix B. Sampling uniformly over the basic probability simplex

We describe two algorithms for sampling uniformly over the basic probability simplex, i.e., $p$ satisfying $p_k \geqslant 0$ and $\sum_k p_k = 1$.

The first algorithm employs the idea that $K$ exponential random variables, when normalized, reduce to a Dirichlet distribution [25],

$$f(x; \alpha) \propto \prod_{k=1}^{K} x_k^{\alpha_k - 1}, \tag{B.1}$$

where the $\alpha$ parameters are positive. Together with the fact that a uniform sample over the basic probability simplex is a Dirichlet distribution with $\alpha_k = 1$ for all $k$, we can sample uniformly over the simplex as follows:

**Bas1** Generate $K$ random numbers $x_1, x_2, ..., x_K$ uniformly over the interval $(0, 1)$.

**Bas2** Set $y_k = -\log(x_k)$. Each $y_k$ obtained this way is drawn from an exponential distribution, i.e., $f(y; \lambda) = \lambda e^{-\lambda y}$, with the rate parameter $\lambda = 1$.

**Bas3** Set $p_k = y_k / \sum_{k=1}^{K} y_k$.

The second algorithm, analyzed in [26], samples as follows:

**Bas'1** Start with $x_0 = 0$ and $x_K = 1$.

**Bas'2** Draw $K - 1$ random numbers uniformly over the interval $(0, 1)$, and sort the list from smallest to largest: $(0 = x_0 \leqslant) x_1 \leqslant x_2 \leqslant \cdots \leqslant x_{K-1} (\leqslant x_K = 1)$.

**Bas'3** Set $p_k = x_k - x_{k-1}$.

For both algorithms, $p = (p_1, p_2, ..., p_K)$ is uniformly distributed over the simplex.

## Appendix C. Probability density for the purity

Prior I of section 4 and figures 2 and 3 is defined by the sampling method employed by Życzkowski *et al* [3]. They choose the eigenvalues $r_1, r_2, ..., r_d$ of the statistical operator $\rho$ uniformly from the simplex with $r_j \geq 0$ and $\sum_j r_j = 1$ and represent the eigenkets of $\rho$ by the eigencolumns of a Haar-random unitary matrix.

For a statistical operator from this sample, the purity is the sum of the squared eigenvalues,

$$\xi(\rho) = \text{tr}\left\{\rho^2\right\} = \sum_{j=1}^{d} r_j^2. \tag{C.1}$$

We write $d\xi q(\xi)$ for the probability that the purity is between $\xi$ and $\xi + d\xi$ and then have

$$q(\xi) = (d-1)! \int_0^\infty dr_1 \int_0^\infty dr_2 \cdots \int_0^\infty dr_d \; \delta\left(\sum_{j=1}^{d} r_j - 1\right) \delta\left(\sum_{j=1}^{d} r_j^2 - \xi\right) \tag{C.2}$$

for $\frac{1}{d} \leq \xi \leq 1$. Explicit expressions for the purity density $q(\xi)$ are

$$d = 2: \quad q(\xi) = \frac{1}{\sqrt{2\xi - 1}} \quad \text{for} \quad \frac{1}{2} \leq \xi \leq 1, \tag{C.3}$$

for the qubit case of figure 3(a);

$$d = 3: \quad q(\xi) = \begin{cases} \dfrac{2\pi}{\sqrt{3}} & \text{for} \quad \dfrac{1}{3} \leq \xi \leq \dfrac{1}{2}, \\[3mm] \sqrt{3}\left[\dfrac{\pi}{6} - \sin^{-1}\left(\dfrac{3\xi - 2}{3\xi - 1}\right)\right] & \text{for} \quad \dfrac{1}{2} \leq \xi \leq 1, \end{cases} \tag{C.4}$$

for the qutrit case of figure 3(b); and

$$d = 4: q(\xi) = \begin{cases} 3\pi\sqrt{4\xi - 1} & \text{for} \quad \dfrac{1}{4} \leq \xi \leq \dfrac{1}{3}, \\[3mm] 2\sqrt{3}\,\pi - 3\pi\sqrt{4\xi - 1} & \text{for} \quad \dfrac{1}{3} \leq \xi \leq \dfrac{1}{2}, \\[3mm] 3\sqrt{3}\left[\cos^{-1}\left(\dfrac{3\xi - 2}{3\xi - 1}\right) - \dfrac{\pi}{3}\right] & \\[3mm] -9\sqrt{4\xi - 1}\left[\sin^{-1}\left(\dfrac{\xi}{3\xi - 1}\right) - \dfrac{\pi}{6}\right] & \text{for} \quad \dfrac{1}{2} \leq \xi \leq 1, \end{cases} \tag{C.5}$$

for the qubit-pair case of figure 2(b).

The prior density of prior II is proportional to $(d\rho)$ of (8) and (7) with $w_0(p) = 1$. In the case of the qubit and the tetrahedron POM of [18], the four probabilities are related to the expectation values of the Pauli operators by equations (25) of [12], and (26) there states the factor $w_{qu}(p)$ for the quantum constraint in (6). The purity is

$$\xi(\rho) = \text{tr}\left\{\rho^2\right\} = 6\sum_{j=1}^{4} p_j^2 - 1, \tag{C.6}$$

in terms of the tetrahedron probabilities. Accordingly, we get the purity density

$$d = 2: q(\xi) = \frac{\int (dp)\; \delta\left(6\sum_{j=1}^{4} p_j^2 - 1 - \xi\right)}{\int (dp)} = 3\sqrt{2\xi - 1} \quad \text{for} \quad \frac{1}{2} \leq \xi \leq 1; \tag{C.7}$$

see figure 3(a). Likewise, we find that

$$d = 3: q(\xi) \propto (3\xi - 1)^3 \quad \text{for} \quad \frac{1}{3} \leq \xi \leq \frac{1}{2},$$
$$d = 4: q(\xi) \propto (4\xi - 1)^{13/2} \quad \text{for} \quad \frac{1}{4} \leq \xi \leq \frac{1}{3}, \tag{C.8}$$

for the qutrit case of figure 3(b) and the qubit-pair case of figure 2(b), but we do not have analytical expressions for larger values of $\xi$ because the quantum constraints introduce complications that are difficult to cope with.

## Appendix D. MHMC algorithm

For a random walk over a parameter space with elements $\theta$ and a target distribution $f(\theta)$, the MHMC algorithm is the following:

**MHMC1** Choose a proposal-generating density, say $J(\theta^*|\theta)$, which describes the probability density of proposing point $\theta^*$ given the current point $\theta$.

**MHMC2** Choose a starting point $\theta^{(1)}$, and set $j = 1$.

**MHMC3** Randomly draw a candidate $\theta^*$ from the density $J(\theta^*|\theta^{(j)})$.

**MHMC4** Compute the acceptance ratio

$$a = \min\left\{\frac{f(\theta^*)J(\theta^{(j)}|\theta^*)}{f(\theta^{(j)})J(\theta^*|\theta^{(j)})}, 1\right\}. \tag{D.1}$$

**MHMC5** Draw a random number $b$ uniformly from the range $0 < b < 1$. If $a > b$, set $\theta^{(j+1)} = \theta^*$; otherwise, set $\theta^{(j+1)} = \theta^{(j)}$. This implements the criterion of accepting the new proposal $\theta^*$ with probability $a$.

**MHMC6** Update $j \to j+1$. Escape the loop when $j=M$, the target number of sample points; otherwise, return to MHMC3.

The proposal-generating density $J$ determines where we move to in the next step of the Markov chain, and it is convenient to choose one that is symmetric, i.e., $J(\theta^*|\theta) = J(\theta|\theta^*)$, for all $\theta$ and $\theta^*$. This symmetry simplifies the expression for the acceptance ratio, and furthermore, relieves us of the need to know the specific form of $J$, apart from enforcing the symmetry. A common symmetric choice for $J(\theta^*|\theta)$ is the (multivariate) Gaussian distribution, with mean $\theta$ and a constant covariance matrix.